\documentclass[10pt,aps.prl,twocolumn,superscriptaddress]{revtex4-1} 

\usepackage{graphics}
\usepackage{amsmath}
\usepackage{float}
\usepackage{subfig}
\usepackage{tabularx}
\usepackage[normalem]{ulem}

\begin{document}
\title{Generalizing the correlated chromophore domain model of reversible photodegradation to include the effects of an applied electric field}
\author{Benjamin Anderson and  Mark G. Kuzyk}
\address{Department of Physics and Astronomy, Washington State University,
Pullman, WA 99164-2814}
\date{\today}

\begin{abstract}
All observations of photodegradation and self healing follow the predictions of the correlated chromophore domain model. [Ramini {\em et.al.} Polym. Chem., 2013, \textbf{4}, 4948.]  In the present work, we generalize the domain model to describe the effects of an electric field by including induced dipole interactions between molecules in a domain by means of a self-consistent field approach.  This electric field correction is added to the statistical mechanical model to calculate the distribution of domains that are central to healing.  Also included in the model are the dynamics due to the formation of an irreversibly damaged species.  As in previous studies, the model with a one-dimensional domain best explains all experimental data of the population as a function of time, temperature, intensity, concentration, and now applied electric field.  Though the nature of a domain is yet to be determined, the fact that only one-dimensional domain models are consistent with observations suggests that they might be made of correlated dye molecules along polymer chains.

\vspace{1em}
OCIS Codes:

\end{abstract}

\maketitle

\vspace{1em}

\section{Introduction}
Organic dyes are widely used in many applications such as liquid dye lasers \cite{Sorokin66.01,Sorokin67.01,Sorokin67.02}, solid state dye lasers (SSDLs)\cite{Soffer67.01,Maslyukov95.01,Duarte03.01,Costela09,Duarte09.01}, organic light emitting diodes (OLEDs)\cite{Tang87.01,Burroughes90.01}, dye sensitized solar cells (DSSCs)\cite{Gerisher68.01,Tributsch73.01,Matsumura80.01,Krunks09.01,Wu12.01,Liang13.01,Hara09.01}, fluorescence microscopy\cite{Lichtman05.01,Dai96.01,Agard89.01,So00.01} and space based optics\cite{taylo05.01,kuzyk07.02,Taylor07.01,Samuel07.01}.  As lasing media, organic dyes offer many advantages over other materials as their broad absorption and emission peaks\cite{kuhn49.01} allow for large tunability and  generation of ultra short pulses\cite{Fork81.01}.  Additionally, organic dyes tend to have very large laser gains making them a highly efficient lasing medium\cite{Duarte90.01}.  In the field of consumer electronics, OLEDs allow for the construction of extremely lightweight, ultrafast, and energy efficient displays.  Similarly, organic dyes in DSSCs offer the possibility of lightweight, inexpensive, efficient solar cells\cite{Gao08.01,ACS06.01,Yella11.01}.  Finally, organic dyes in fluorescence microscopy allow for high resolution confocal imaging\cite{Dai96.01}, three dimensional imaging\cite{Agard89.01} and two-photon imaging\cite{So00.01}.

While organic dyes offer many benefits over other materials, one fundamental hurdle is their photostability.  Once degraded the dyes must be replaced, which in many applications is impractical, costly and hazardous.  To address the photostability of organic dyes, extensive work has focused on understanding and limiting the effects of photodegradation\cite{wood03.01,taylo05.01,exarh98.01,White91.01,Rabek95.01,Cerdan12.01,Yunus04.01,Fellows05.01,Kurian02.01,Zhang98.01,Mortazavi94.01,Sutherland96.01,Annieta,Tanaka06.01, Albini82.01,Dubois96.01,Rahn94.01,Avnir84.01,Knobbe90.01,Kaminow72.01,Tributsch73.01,Matsumura80.01,Krunks09.01,chu05.01,norwo90.02,Wang95.02,vigil98.01}.  One method found to mitigate photodegradation of a dye is to dope it into a solid matrix such as a sol-gel \cite{Dubois96.01,Rahn94.01}, silicate gel \cite{Avnir84.01,Knobbe90.01} or polymer \cite{Kaminow72.01,howel02.01,howel04.01}.  Remarkably, in the case of some dye-doped polymers, photodegradation is not only mitigated, but is found to be completely reversible\cite{howel02.01,howel04.01,embaye08.01,zhu07.01,zhu07.02}.

Reversible photodegradation is a relatively new phenomenon with the first reported example being fluorescence decay and recovery of Rhodamine and Pyrromethene dye-doped polymer optical fibers\cite{peng98.01}.  Several years later the anthraquinone derivative 1-Amino-2-methylanthraquinone (disperse orange 11 or DO11) doped into (poly)methyl-methacrylate (PMMA) was found to decay reversibly as probed by amplified spontaneous emission (ASE)\cite{howel02.01,howel04.01,embaye08.01}.  Other anthraquinone derivatives have also been shown to exhibit self healing\cite{Anderson11.02}, as well as 8-hydroxyquinoline aluminum (Alq)\cite{Kobrin04.01} and the octopolar molecule AF455\cite{zhu07.01,zhu07.02,Desau09.01}.

Photodegradation is generally an irreversible light-driven chemical reaction that produces new species, such as fragments of the original molecule.  The formation of a new species is characterized by a change in the UV-VIS absorption spectrum.  In a two component system in which molecules of one species (undamaged) are converted to a different (damaged) species, all spectra cross at an isosbestic point.  DO11 molecules, the focus of our present studies, are found to photodegrade irreversibly in {\em liquid methyl-methacrylate} (MMA) monomer -- as is typical for molecules in solution -- with an isosbestic point in its UV-VIS absorption spectrum.  In the {\em solid polymerized state} of MMA, i.e. PMMA polymer, the same experimental conditions show photodegradation with a similar isosbestic point\cite{Ramini12.01}, but the material subsequently recovers when the pump light is turned off.  Thus, the photodegradation pathway of the reversible process, as characterized by UV-VIS spectroscopy, appears to be of the same type as the irreversible process.  Despite the term ``photodegradation'' typically implying irreversibility, since the decay pathway appears to be the same in both solution and polymer we still call the reversible process {\em photodegradation}.

An argument may be made that the more complex structure of a polymer, with a distribution of chain lengths and inhomogeneity in rheological properties and composition, leads to transient processes that may mimic photodegradation and healing, but are in fact something different.  However, since the spectra of what would be characterized as true photodegradation in liquid samples and in PMMA are similar suggests that the origins of the degradation processes are related, but with degradation in PMMA also exhibiting a recovery process.  

There are several possible processes that could mimic self-healing, with the dye molecules still remaining damaged, including: recovery due to polymer relaxation, orientational hole burning, and photo thermally-induced diffusion of dye molecules.  However, we find experimental evidence discounting these mechanisms as valid explanations.  In the case of recovery processes that rely on polymer relaxation, they are known to be accelerated at elevated temperature.\cite{singe91.01}  In contrast, the self-healing process slows at elevated temperature.\cite{Ramini12.01,Ramini13.01}.  In the case of orientational hole burning, there should be a measurable change in linear dichroism due to decay and recovery, but experimentally there is no such observation\cite{embaye08.01}. For photothermally-induced diffusion of dye molecules, the UV-VIS absorbance spectrum will not have an isosbestic point and the burned area should show visible signs of diffusion.  Yet for reversible photodegradation we do observe an isosbestic point and direct imaging of a burn line reveals the process to be inconsistent with diffusion.\cite{Ramini11.01} In addition, the observed temperature dependence of healing is opposite to what would be expected for diffusion.

Self healing, while observed in many systems, is not universal and seems to rely on interactions between particular combination of polymers and dopants.  Given the evidence, it may be a new phenomena that is not derivable from the same old suspects.  In an effort to understands it's origin, many experiments and theoretical models have been developed.

For example, Embaye \textit{et.al.} developed a simple two-species model\cite{embaye08.01} in which a pristine molecule is damaged by photoinduced tautomerization that forms semi-stable dimers rather than non-reacting molecular fragments.  Recovery was proposed to be in the dissociation of dimers back into single molecules.  While Embaye's two-species model fits most experimental data at fixed temperature, concentration, and applied electric field, the theory's parameters do not predict the temperature, concentration, and applied electric field dependence, so does not lead to any insights into the mechanism of healing.

To model temperature- and concentration-dependent reversible photodegradation of ASE in DO11/PMMA, Ramini \textit{et.al.} developed a correlated chromophore domain model (CCDM)\cite{Ramini12.01} in which the dye molecules form correlated domains that foster the interaction between molecules, which promotes self healing.  From experimental data the domains are found to be isodesmic (binding energy independent of domain size), which is typically only true for linear arrays of molecules\cite{Duque97.01,Collings10.01,Cates90.01,McKitterick10.01,Coopersmith63.01,Henderson09.01,Maiti02.01}. The CCDM model assumes that healing is mediated by neighboring molecules, so the recovery rate increases in proportion to the domain size.  At higher temperature, the average domain size decreases, inhibiting healing, thus explaining the observed decrease in recovery rate.

The CCDM was based on determining the population of undamaged molecules with ASE experiments as a function of temperature and concentration, but due to experimental constraints could only be tested for a limited range of fluences.  Simulations using the CCDM predicts a strong dependence of the recovery rate on the fluence.  However, linear transmittance measurements with fluences ranging over several orders of magnitude find the recovery rate to be constant as a function of fluence\cite{Anderson11.01,Anderson11.02}.  To reconcile the CCDM with this data, the CCDM's recovery dynamics were modified to include the effects of the damaged species in a domain and found to be in agreement -- within a constant offset -- with linear transmittance measurements\cite{Ramini13.01}.  The constant offset \cite{Anderson11.01,Anderson11.02,Anderson12.01,Anderson12.03,Anderson13.01} is attributed to an irreversibly damaged species not measurable by ASE.  In contrast, transmittance is sensitive to both.

While the latest CCDM model\cite{Ramini13.01} predicts the behavior of a broad range of experimental data, the nature of a domain remains elusive.  The concept of the domain as the critical ingredient to the recovery process was introduced into the model because it works.  Should domains not be responsible for the healing process, the mathematical structure of the correct theory would necessarily be the same.

Validation of a theory built on data from a given space of parameters requires the theory to be extended to an orthogonal space, and the predictions of the generalized theory to be tested in the new space.  In this paper, we extend the model to include the effects of an electric field on a domain and test the model's predictions with electric-field-dependent experiments.

Recent measurements have shown that an applied electric field affects the photodegradation and recovery process.\cite{Anderson12.01,Anderson12.03,Anderson13.01}  In particular,
applying a constant electric field during photodegradation and recovery has been shown to
\begin{enumerate}
\item Decrease the decay rate.
\item Decrease the amount of damage.
\item Decrease the recovery rate.
\item Increase the recovery fraction.
\end{enumerate}

The purpose of this work is to extend the CCDM using a self-consistent local field model of molecular interactions to take into account the effect of an electric field on domain size; and, to test the predictions of this domain-based model on the new observations.  Since the measurements are of samples exposed to high doses of light, an irreversible decay product is also formed.  To take this into account, we include the effects of an irreversibly damaged species in the model.  This generalized model is found to predict the observed effect of an electric field on reversible photodegradation.  As such, we will see that the domain continues to be the common factor in all models that predict the observed behavior.

\section{Model}

\subsection{Extension of domain model to include three species}
The domain model of reversible photodegradation was initially developed using data obtained from ASE measurements in which only two species appear to be present, an undamaged species and a reversibly damaged species\cite{Ramini12.01,Ramini12.02,Kuzyk12.01,Kuzyk12.03,Ramini13.01}.  However, measurements using linear optical transmittance techniques, such as transmittance imaging and absorbance spectroscopy, have shown that during decay a third species is formed which is an irreversible product of optical damage\cite{Anderson11.01,Anderson11.02,Anderson12.01,Anderson12.03,Anderson13.01}.

To explain the irreversible species, which linear measurements observe but nonlinear measurements do not, we consider the nature of the two types of measurements. Nonlinear measurements  primarily measure the dye's optical properties, as PMMA has a negligible nonlinear susceptibility.  Linear optical measurements, on the other hand,  measure both the dye's and polymer's optical properties.  We therefore hypothesize that the irreversibly damaged species is due to photodamage of the polymer, with the reversibly damaged species being related to the dye.

\begin{figure}
\centering
\includegraphics{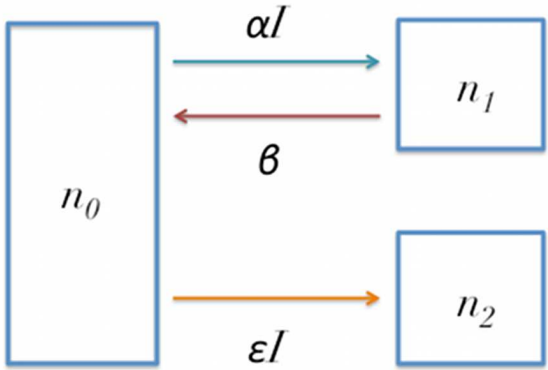}
\caption{Schematic diagram of the three-species process.  The decay processes occur in parallel, but only one species recovers.}
\label{Fig;3states}
\end{figure}

We assume that irreversible decay occurs simultaneously with reversible decay, i.e. the decay channels are parallel.  A schematic representation of this process is shown in Figure \ref{Fig;3states}, where the undamaged molecules absorb light and either decay into the reversible species, or the irreversible species. Mathematically, this system is modeled by three rate equations,

\begin{eqnarray}
\frac{dn_0}{dt}&=&-\left(\frac{\alpha}{N}+\epsilon N\right) I n_0+\beta N n_1, \label{Eqn:n0rate}
\\ \frac{dn_1}{dt}&=&\frac{\alpha I}{N}n_0-\beta N n_1,
\\ \frac{dn_2}{dt}&=&\epsilon N I n_0,  \label{Eqn:n2rate}
\end{eqnarray}
where $N$ is the domain size, $\alpha$ is the intensity independent decay rate of the reversible decay process, $\epsilon$ is the intensity independent decay rate of the irreversible process, $\beta$ is the recovery rate, and $I$ is the intensity. The domain size dependence of both the reversible decay rate ($\frac{\alpha}{N}$) and the recovery rate ($\beta N$) are retained from Ramini's CCDM\cite{Ramini13.01}, and the domain size dependence of the irreversible decay rate is chosen to match experimental observations with linear transmittance imaging\cite{Anderson11.01,Anderson11.02, Anderson12.01,Anderson12.02}.  We hypothesize that the domain size dependence of irreversible degradation is due to polymer damage.  In experiments comparing undoped PMMA to dye doped PMMA, we have found that there is no appreciable change in absorbance spectrum for the undoped PMMA under our standard experimental conditions.  However, once the polymer is doped with dye we find that an irreversible species begins to occur.

There are several possible mechanisms for irreversible degradation of the polymer host.  One is that light absorption by the dye leads to photo thermal heating of the polymer, resulting in irreversible changes to the polymer structure.  Another possibility is that light absorbed by dyes produces free ions/radicals which interact with the polymer chain forming either singlet oxygen\cite{Rabek95.01,Sutherland96.01,Kurian02.01,Cerdan12.01} and/or electron donor-acceptor complexes\cite{Kochi91.01,Annieta}.  The singlet oxygen and/or electron donor-acceptor complexes can then produce irreversible chemical reactions including hydrogen abstraction\cite{Rabek95.01,Tanaka06.01, Albini82.01}, carbonyl group formation \cite{Cumpston95.01,Sutherland96.01}, hydroperoxide formation\cite{Li84.01}, and/or other reactions\cite{Rabek95.01}. Independent of the mechanism of polymer damage, it will become greater as the dye concentration is increased, which is consistent with the irreversible intensity independent decay rate scaling as $\epsilon N$.

To solve for the population dynamics, Equations \ref{Eqn:n0rate}-\ref{Eqn:n2rate} are written in matrix form as
\begin{equation}
\frac{d\mathbf{x}}{dt}=\mathbf{\xi}\mathbf{x},
\label{eqn:vecdiff}
\end{equation}
where the column vector $\mathbf{x}=(n_1,n_2,n_3)$ and the matrix $\mathbf{\xi}$ is

\begin{equation}
\mathbf{\xi}=\left(
\begin{array}{ccc}
 -(\frac{\alpha}{N}+\epsilon N) I & \beta N  & 0 \\
\frac{\alpha}{N} I & -\beta N  & 0 \\
 \epsilon N I & 0 & 0
\end{array}
\right).
\end{equation}
The general solution to Equation \ref{eqn:vecdiff} is
\begin{equation}
\mathbf{x}=c_0\mathbf{v}_0e^{\lambda_0 t}+c_1\mathbf{v}_1e^{\lambda_1 t}+c_2\mathbf{v}_2e^{\lambda_2 t},
\label{eqn:vecsol}
\end{equation}
where $\lambda_i$ are the eigenvalues of $\mathbf{\xi}$, $\mathbf{v}_i$ are the eigenvectors, and $c_i$ are constants found from the initial conditions.  To simplify the form of the eigenvalues and eigenvectors we define three parameters:
\begin{eqnarray}
A=\beta N +\left(\frac{\alpha}{N} +\epsilon N \right) I,
\\ C=\beta N +\left(\frac{\alpha}{N} -\epsilon N \right) I,
\\ B=\sqrt{-4 \beta  \epsilon N^2   I+(\beta N +\left(\frac{\alpha}{ N} +\epsilon N\right ) I)^2}.
\end{eqnarray}
The eigenvalues are then given by,

\begin{eqnarray}
\lambda_0&=&0,
\\ \lambda_1&=&\frac{-A-B}{2},
\\ \lambda_2&=&\frac{-A+B}{2},
\end{eqnarray}
and the eigenvectors are

\begin{align}
\mathbf{v}_0=\left(
\begin{array}{c}
0
\\  0
\\ 1
\end{array}
\right) \quad
\mathbf{v}_1=\frac{1}{2\epsilon N I}\left(
\begin{array}{c}
-A-B
\\  C+B
\\ 2 \epsilon N I
\end{array}
\right)
\\ \mathbf{v}_2=\frac{1}{2\epsilon N I}\left(
\begin{array}{c}
-A+B
\\  C-B
\\ 2\epsilon N I
\end{array}
\right).
\end{align}

Assuming the system is initially undamaged, the population dynamics of a domain of size $N$ are

\begin{align}
n_0(t)&=\frac{1}{2 B (A-C)}e^{-\frac{1}{2} (A+B) t} \bigg((A+B) (B-C) \nonumber
\\ &+(A-B) (B+C) e^{B t}\bigg), \label{eqn:n0}
\\[1em] n_1(t)&=\frac{(B-C) (B+C) e^{-\frac{1}{2} (A+B) t} \left(-1+e^{B t}\right)}{2 B (A-C)},  \label{eqn:n1}
\end{align}
and
\begin{align}
n_2(t)&=-\frac{\epsilon N I}{B (A-C)}e^{-\frac{1}{2} (A+B) t} \bigg[C \left(-1+e^{B t}\right)+ \nonumber
\\ &B \left(1+e^{B t}-2 e^{\frac{1}{2} (A+B) t}\right)\bigg].  \label{eqn:n2}
\end{align}

Equations \ref{eqn:n0}-\ref{eqn:n2} are the population dynamics for a single domain, where the macroscopic dynamics are an ensemble average over a distribution of domains size $N$, $\Omega(N)$.  Using an isodesmic aggregation model, Ramini \textit{et.al.} derived the distribution of domains of size $N$ to be\cite{Ramini12.01,Ramini13.01}:

\begin{equation}
\Omega(N)=\frac{1}{z}\left[\frac{(1+2\rho z)-\sqrt{1+4\rho z}}{2\rho z}\right]^N.
\label{eqn:domdist}
\end{equation}
where $\Omega(N)$ is the distribution function, $z=\exp(\mu/kT)$, $\rho$ is the total number of molecules in a given volume, and $\mu$ is the free energy advantage of having a molecule being in a domain versus outside of a domain.

The free energy advantage is found by comparing the energy of a domain size $N$ to the energy of a free molecule and a domain of size $N-1$,

\begin{equation}
\mu=E(N)-(E(N-1)+E(1)), \label{eqn:mudef}
\end{equation}
where $E(N)$ is the energy of a domain of size $N$, and $E(N-1)+E(1)$ is the energy of a domain of size $N-1$ and a single molecule outside of the domain.

\subsection{Effect of an electric field on the distribution of domains}
To model the effect of an applied electric field, $E_0$, on the distribution of domains, we consider the change in free energy advantage due to an applied electric field.

\begin{figure}
\centering
\includegraphics{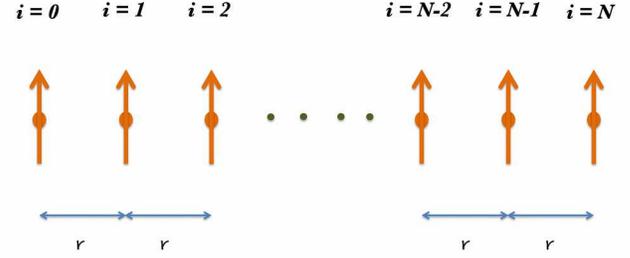}
\caption{Diagram of model system geometry of $N$ equally-spaced point dipole molecules.}
\label{fig:dipoles}
\end{figure}

Our dielectric model, which is an extension of Ramini's model, assumes that a domain is a linear array of equally spaced point dipoles each having polarizability $\alpha$, as shown in  Figure \ref{fig:dipoles}.

In the dilute case, the molecules are essentially noninteracting, and the total dipole moment of the domain is

\begin{align}
P&=\sum_{i=1}^N p_i,
\\&=N\alpha E_0,
\end{align}
where $N$ is the size of the domain and $p_i$ is the dipole moment of the $i^{th}$ molecule in the domain.  Additionally, the total dielectric energy of the domain is

\begin{align}
U(N) &=-N\alpha E_L^2 \label{eqn:ELE}
\\ &=-N\alpha E_0^2  \label{eqn:nointE}
\end{align}
where $E_L=LE_0$ is the local electric field with $L=1$ being the local field factor for the noninteracting case.  Substituting Equation \ref{eqn:nointE} into Equation \ref{eqn:mudef} yields $\mu=0$, thus in the noninteracting case there is no change in the distribution of domains due to the application of an applied electric field.

To account for dielectric interactions between molecules in a domain, we assume that  each molecule behaves as a point dipole, and the field from all other dipoles contribute to the local field .  To solve this system we use a self-consistent field model, similar to Dawson \textit{et.al.} \cite{Dawson11.01,Dawson11.02}.  Assuming that the interactions occur only between molecules in the same domain, we can write the dipole moment of the $i^{th}$ molecule in a domain as

\begin{equation}
p_i=\alpha\left[E_0-\sum_{j=1}^{i-1}\frac{p_j}{((i-j)r)^3}-\sum_{j=i+1}^{N}\frac{p_j}{((j-i)r)^3}\right],
\label{eqn:pi}
\end{equation}
where the effect of the other molecules in the domain is to decrease the local field experienced by the $i^{th}$ molecule.  The total dipole moment of the domain is found by summing over the dipole moments of every molecule in the domain,

\begin{align}
P&=\sum_{i=1}^N p_i,
\\&=N\alpha E_0-\sum_{i=1}^N\sum_{j\neq i}^N\frac{p_j}{(|i-j|r)^3}.
\end{align}

To find the individual dipole moments we can rewrite Equation \ref{eqn:pi} as a matrix equation

\begin{equation}
\mathbf{P}=\alpha E_0\mathbf{1}-\frac{\alpha}{r^3}\mathbf{M}\mathbf{P},
\end{equation}
where the column vector $\mathbf{P}=\{p_1,p_2,\cdots,p_N\}$, the column vector $\mathbf{1}=\{1,1,\cdots,1\}$ and $\mathbf{M}$ is an $N \times N$ matrix with elements given by,

\begin{equation}
 M_{ij} =
  \begin{cases}
   0  & \text{if } i=j \\
   \frac{1}{|i-j|^3}       & \text{if } i \neq j
  \end{cases}.
\end{equation}
Solving for $\mathbf{P}$ we obtain

\begin{equation}
\mathbf{P}=\alpha E_0 \left(\mathbf{I}+\frac{\alpha}{r^3}\mathbf{M}\right)^{-1}\mathbf{1},
\label{eqn:pivec}
\end{equation}
where $\mathbf{I}$ is the identity matrix, and the superscript $-1$ denotes the matrix inverse.

It is beyond the scope of this paper to present a comprehensive solution to Equation \ref{eqn:pivec}. Instead, we consider the total dipole moment of a domain in the case where $\alpha<<r^3$, which is given by

\begin{align}
P(N)&\approx N\alpha E_0-2\zeta(3)(N-1)\frac{\alpha^2}{r^3}E_0,
\\ & = N\alpha L(N) E_0,  \label{eqn:domdipole}
\end{align}
where $\zeta$ is the zeta function with $\zeta(3) \approx 1.202$, and $L$ is the local field factor given by

\begin{equation}
L(N)=1-\frac{2\zeta(3)(N-1)}{N}\frac{\alpha}{r^3}.  \label{eqn:LF}
\end{equation}
Substituting Equation \ref{eqn:LF} into Equation \ref{eqn:ELE} the dielectric energy for a domain of size $N$ including first-order interactions is

\begin{align}
U(N)&=-N\alpha L(N)^2E_0^2,
\\& \approx -N\alpha E_0^2+4\zeta(3)(N-1)\frac{\alpha^2}{r^3}E_0^2,
\end{align}
where terms proportional to $\alpha^2/r^6$ are again neglected.

With the dielectric energy of a domain size $N$ and the aggregation energy used by Ramini \textit{et.al}, $E=-\lambda(N-1)$\cite{Ramini12.01,Ramini13.01}, the total energy of a domain is:

\begin{align}
E(N)=-\lambda(N-1)-N\alpha E_0^2+4\zeta(3)(N-1)\frac{\alpha^2}{r^3}E_0^2.   \label{eqn:DE}
\end{align}
 Substituting Equation \ref{eqn:DE} into Equation \ref{eqn:mudef} we find the free energy advantage, including dielectric effects, to be:
\begin{equation}
\mu=\lambda-\frac{4\zeta(3)\alpha^2}{r^3}E_0^2.
\end{equation}
Therefore with an applied electric field the $z$ parameter in Equation \ref{eqn:domdist} becomes

\begin{align}
z&=\exp\left\{\frac{\mu}{kT}\right\}
\\ &=\exp\left\{\frac{\lambda}{kT}-\frac{4\zeta(3)\alpha^2}{kTr^3}E_0^2\right\}
\\ &=\exp\left\{\gamma-\eta E_0^2\right\}
\end{align}
where $\gamma=\frac{\lambda}{kT}$ and $\eta=\frac{4\zeta(3)\alpha^2}{kTr^3}$.

\subsection{Integrated Model}

Linear optical measurements do not directly measure the three species.  Instead they probe a linear combination weighted by spectral properties of each species.   In the approximation of a thin sample  -- such that the pump intensity is constant as a function of depth -- the absorbance, $A$, may be written as

\begin{equation}
A=\overline{n}_0(t)\sigma_0L+\overline{n}_1(t)\sigma_1L+\overline{n}_2(t)\sigma_2L,
\end{equation}
 where $L$ is the sample thickness, $\sigma_i$ is the absorbance cross section for the $i^{th}$ species, and the populations $\overline{n}_i$ are the ensemble average over the distribution of domains given by

\begin{eqnarray}
\overline{n}_0(t)=\sum^{\infty}_{N=1}n_0(N,t)\Omega(N),  \label{eqn:n0int}
\\ \overline{n}_1(t)=\sum^{\infty}_{N=1}n_1(N,t)\Omega(N),
\\ \overline{n}_2(t)=\sum^{\infty}_{N=1}n_2(N,t)\Omega(N), \label{eqn:n2int}
\end{eqnarray}
where $n_i(N,t)$ are given by Equations \ref{eqn:n0}-\ref{eqn:n2} and the distribution of domains, $\Omega(N)$, is given by Equation \ref{eqn:domdist}.

Assuming that the region of interest is originally undamaged, the change in absorbance due to photodegradation is

\begin{align}
\Delta A&=\overline{n}_1(t)(\sigma_1-\sigma_0)L+\overline{n}_2(t)(\sigma_2-\sigma_0)L,
\\ &=\overline{n}_1(t) \Delta \sigma_1L+\overline{n}_2(t) \Delta \sigma_2L
\end{align}
where $ \Delta \sigma_i=\sigma_i-\sigma_0$. For consistency with our previous papers\cite{Anderson11.01,Anderson11.02,Anderson12.01,Anderson12.03,Anderson13.01} we label $\Delta A$ as the scaled damaged population, $n'$, which is proportional to the damaged population.

\section{Results}
To test the extended correlated chromophore domain model (eCCDM) we use data from our previous study of electric field dependent reversible photodegradation\cite{Anderson13.01}.  The data set consists of the scaled damaged population at the burn center during decay and recovery, as well as the reversible and irreversible amplitudes found at a wide range of intensities.  Each measurement is repeated for five different applied electric field strengths with the polarity of the applied field held constant throughout all testing.  The theory is fit to the full data set with one adjustable parameter, which accounts for point-to-point variations in the sample.

\begin{figure}
\centering
\includegraphics{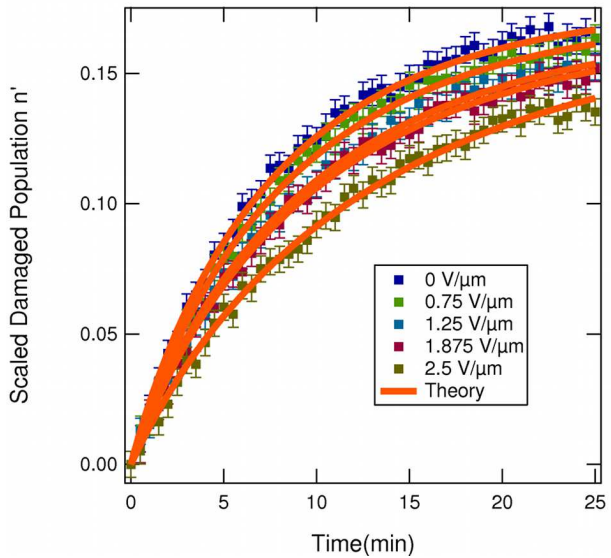}
\caption{Decay of scaled damaged population as a function of time (points) for different applied fields for an intensity of 175W/cm$^2$, and theory (curves).}
\label{Fig:dec}
\end{figure}

\begin{figure}
\centering
\includegraphics{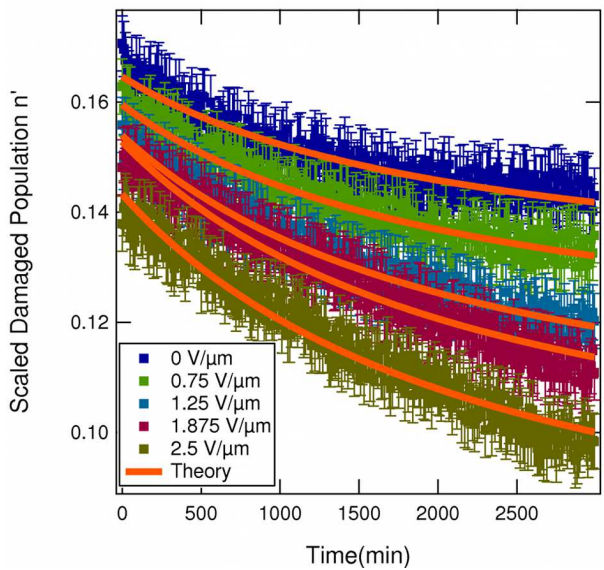}
\caption{Recovery of scaled damaged population as a function of time (points) for different applied fields after a 25 min burn of intensity of 175W/cm$^2$, and theory (curves).}
\label{Fig:rec}
\end{figure}

\begin{figure}
\centering
\includegraphics{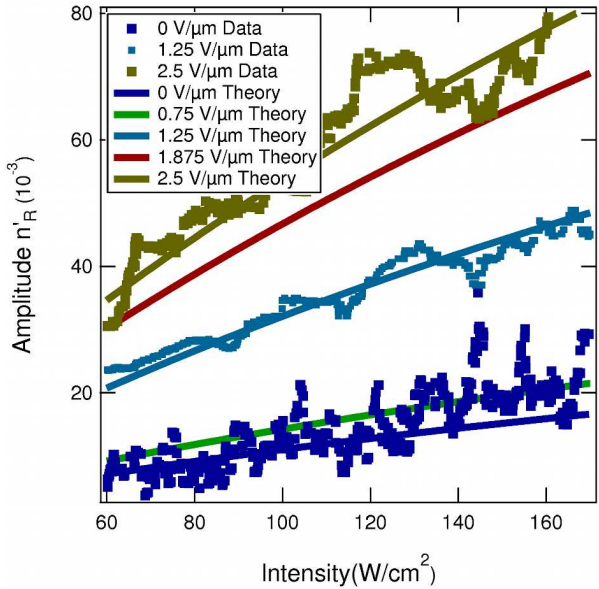}
\caption{Measured amplitude of recovered population as a function of intensity (points) and theory (curves).  The amplitude scales with the reversibly damaged population $n_1$ (Equation \ref{eqn:n1}).  Fits for each field strength are displayed but smoothed data is shown for only three representative field strengths to avoid clutter.}
\label{Fig:Amp}
\end{figure}

\begin{figure}
\centering
\includegraphics{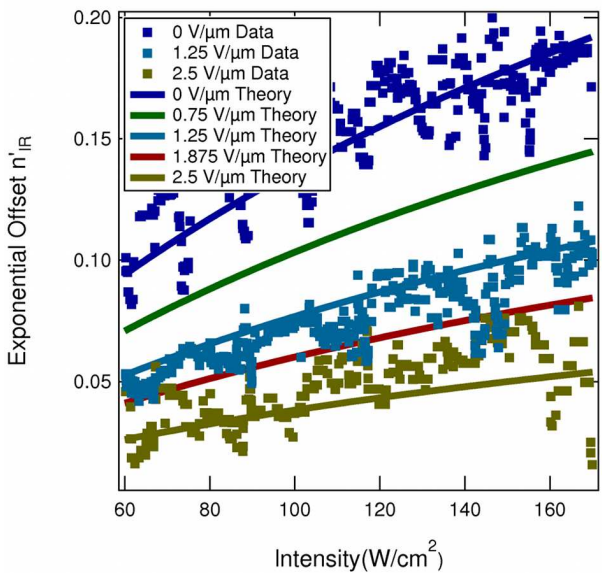}
\caption{Offset ($n'_{IR}$) in recovery data. Plots of the offset as a function of intensity are shown (points), with theory (curves).  Population of irreversibly damaged species, $n_2$ (Equation \ref{eqn:n2}), formed during decay is manifested in an offset $n_{IR}$.   Fits for each field strength are displayed but smoothed data is shown for only three representative field strengths to avoid clutter.}
\label{Fig:off}
\end{figure}

Figures \ref{Fig:dec} and \ref{Fig:rec} show the scaled damage population decay and recovery, respectively, for the pump beam center, which has an intensity of 175 W/cm$^2$.  Using points along the pump beam profile --as samples of different intensity -- we fit the scaled damage population during recovery as a function of time using an exponential fit,

\begin{equation}
n'=n'_{IR}+n'_{R}e^{-\beta t},
\end{equation}
where the exponential offset, $n_{IR}$, is proportional to the irreversibly damaged population formed during photodegradation, and the exponential amplitude, $n_{R}$, is proportional to the reversibly damaged population remaining right after degradation.  Figure \ref{Fig:Amp} shows the exponential amplitude as a function of intensity, and Figure \ref{Fig:off} shows the exponential offset as a function of intensity. While  the full data set is used for fitting, Figure \ref{Fig:Amp} and \ref{Fig:off} only show smoothed data for three field strengths as the raw data is extremely noisy due to point-to-point variations due to sample inhomogeneity.

The eCCDM is fit to the full data set simultaneously with the model parameters held consistent, with one adjustable parameter used to account for point-to-point variations in the sample, leading to a well defined set of parameters as tabulated in Table \ref{Tab:Parameters}.   To compare the results of the eCCDM to the previous CCDM we include the domain parameters ($\rho$ and $\lambda$) found previously using ASE as probe\cite{Ramini12.01,Ramini13.01}.  However, given the differences in the models we are unable to directly compare the rate parameters.

\begin{table}[H]
\begin{center}
\begin{tabular}{|c|cc|}
\hline
Parameter   & New  & Old \\[1em]  \hline
$\alpha$(10$^{-2}$ cm$^2$/(W min) ) &  $1.32 (\pm 0.33)$ \qquad& -\\[1em]
$\beta$(10$^{-5}$ min$^{-1}$) &  $2.53 (\pm  0.51)$ \qquad& - \\[1em]
$\epsilon$(10$^{-6}$ cm$^2$/(W min) ) & $6.47 (\pm 0.21$) \qquad & - \\[1em]
$\rho$(10$^{-2}$) &$ 1.19 (\pm 0.25 $)  \qquad& $ 1.2 (\pm 0.2) $ \\ [1em]
$\lambda$(eV) & $0.29 (\pm 0.02$) \qquad& $0.29 (\pm 0.01)$ \\[1em]
$\eta$(10$^{-14}$ m$^2$V$^{-2}$ )  & $2.210 (\pm  0.070$)  \qquad& - \\ \hline
\end{tabular}
\end{center}
\caption{Model parameters for the eCCDM found from electric field dependent reversible photodegradation measurements.  The thermodynamic quantities, $\rho$ and $\lambda$, are compared to previous measurements\cite{Ramini12.01,Ramini13.01}}
\label{Tab:Parameters}
\end{table}

\section{Discussion}
\begin{figure}
\centering
\includegraphics{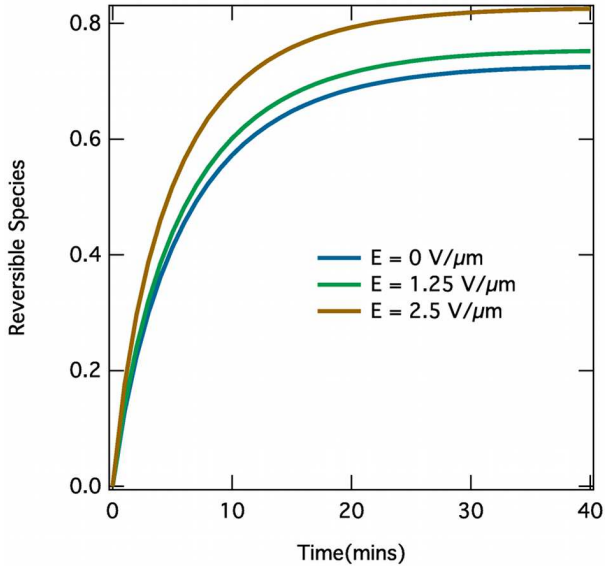}
\caption{Reversibly damaged species as a function of time for three applied field strengths during decay.}
\label{fig:RDCE}
\end{figure}

\begin{figure}
\centering
\includegraphics{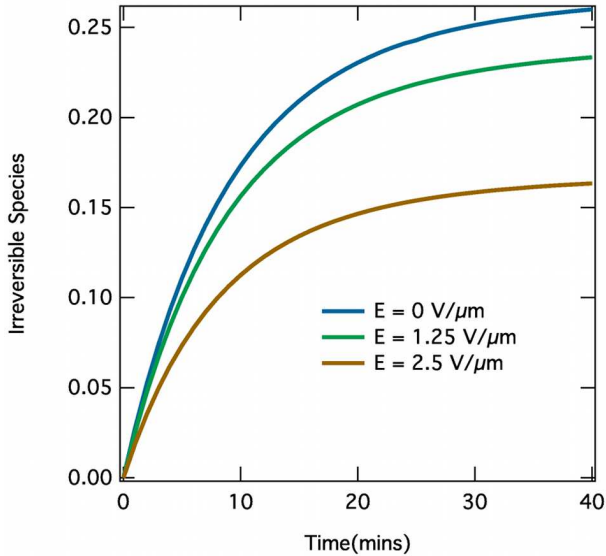}
\caption{Irreversibly damaged species as a function of time for three applied field strengths during decay.}
\label{fig:IDCE}
\end{figure}

The previous section shows that the eCCDM accurately describes transmittance imaging data as a function of time, intensity and applied electric field. However, the underlying population dynamics are masked, as the scaled damaged population is a linear superposition of both damaged species.   Therefore, using the model parameters in Table \ref{Tab:Parameters}, we determine the underlying population dynamics.  Figure \ref{fig:RDCE} shows the reversibly decayed species as a function of time during decay for three field strengths, and Figure \ref{fig:IDCE} shows the irreversibly decayed species during decay for three field strengths.

As the applied field is increased, more of reversibly damaged population is produced and at a faster rate.  On the other hand, with increasing field strength, less of the irreversibly damaged population is formed and at a slower rate.  This observation is opposite of previous measurements of purely irreversible decay of dye-doped polymers under an applied electric field\cite{Khan97.01,Su13.01,Kang99.01}.  These results, coupled with previous measurements of temperature dependent reversible photodegradation\cite{Ramini12.01,Ramini13.01} suggest that the underlying mechanism of reversible photodegradation is unique.

A domain model assuming linear aggregates of correlated dye molecules fits all experimental data as a function of intensity, temperature, concentration and applied electric field.  In this model the dynamics of decay and recovery are governed by the distribution of domain sizes.  A precise calculation of the effect of varying concentration, temperature and/or electric field on population dynamics requires calculating the ensemble average in Equations \ref{eqn:n0int}-\ref{eqn:n2int}.  A more simple, but approximate, way to glean an understanding of the dynamics is to consider the average domain size as a function of temperature, concentration and applied electric field.

\begin{figure}
\centering
\includegraphics{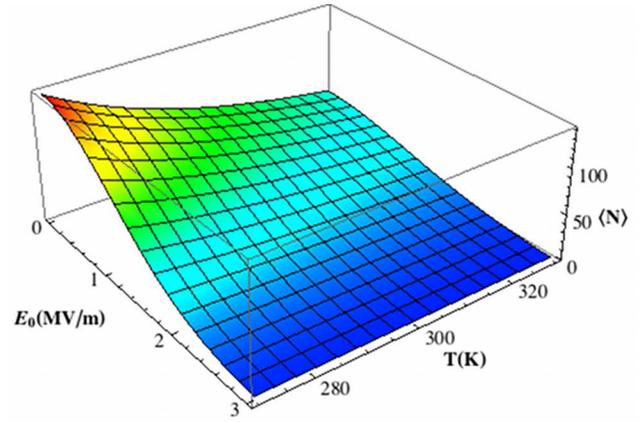}
\caption{Average domain size as a function of temperature and applied electric field using parameters found from fits.}
\label{fig:prob}
\end{figure}

The probability of a domain having size $N$ is $P(N)=N\Omega(N)$.  Therefore the average domain size is:

\begin{align}
\langle N \rangle&=\frac{\sum_{N=1}^{\infty}NP(N)}{\sum_{N=1}^{\infty}P(N)},
\\ &=\frac{1}{\rho z}\sum _{N=1}^{\infty } N^2\left(z\Omega _1\right)^N,
\\ &=\frac{\Omega _1 \left(1+z \Omega _1\right)}{\rho  \left|z \Omega _1-1\right|^3},
\end{align}
where $z=\exp\{\mu/kT\}$, $\rho=\sum_{N=1}^{\infty}P(N)$, and

\begin{equation}
\Omega_1=\frac{(1+2\rho z)-\sqrt{1+4\rho z}}{2\rho z^2}.
\end{equation}
Figure \ref{fig:prob} shows the average domain size as a function of temperature and applied electric field for $\rho=0.012$.  As the temperature and/or applied field is increased the average domain size decreases.  This implies that the effect of an applied field and/or temperature increase is to break apart domains into smaller sizes.
To understand this effect we recall that the free energy advantage $\mu$, is essentially  a binding energy describing the attachment of molecules to a domain.  When increasing the applied electric field, the dipole-dipole interactions weaken the overall binding energy making it easier for a molecule to break free of a domain.  Additionally, when increasing the temperature, the greater thermal energy causes domains to break apart.

Despite the model's success, the nature of domains is still unknown. Based on the following experimental/modeling evidence:

\begin{enumerate}
\item A dye must be in a polymer matrix to exhibit self healing\cite{howel02.01,howel04.01}.
\item Not only does the polymer help the dye recover, but dye can help the polymer recover \cite{Desau09.01}.
\item The irreversibly decayed species appears to be related to polymer damage, suggesting a close correlation between dye and polymer.
\item Domain geometries, other than linear aggregation, have been tested and fail to fit experimental data.  Since polymers are essentially one dimensional chains, we speculate that molecules may form linear aggregates along a chain.
\end{enumerate}
We propose that domains consist of molecules correlated with each other through a single polymer chain. Currently the nature of how dyes and polymer chains form domains is under study, but given the measured domain binding energy of $\lambda=0.29$eV, Kuzyk \textit{et.al} proposed simple hydrogen bonding scheme between a DO11 tautomer and PMMA which reproduces the binding energy\cite{Kuzyk12.01,Kuzyk12.03}.  To further test this hypothesis studies are underway that vary dye structure and polymer type, as these will change the binding energy if hydrogen bonding between the dye and polymer is responsible for domain formation.  Additionally, new experimental methods such as FTIR, micro-Raman, and scattering are being pursued in order to better understand the nature of domains.

\section{Conclusions}
Using a dielectric model of a linear array of equally spaced dipoles, we have extended the correlated chromophore domain model to account for the effects of an applied electric field on reversible photodegradation.  The new model is observed to be consistent with all experiments in which an electric field is applied.  As such, the domain appears to be the unifying factor.

In both the case of the aggregation energy, $\lambda$, and the electric field induced energy, the free energy advantage is found to be independent of domain size, suggesting linear aggregates\cite{Duque97.01,Collings10.01,Cates90.01,McKitterick10.01,Coopersmith63.01,Henderson09.01,Maiti02.01}.  Most likely these aggregates are molecules correlated with each other through a polymer chain. This picture is consistent with the hypothesis that the irreversible species is due to photo-induced damage to the polymer that is mediated by energy transfer from a dye upon photodegradation.

\section{Acknowledgments}
We would like to thank Sheng-Ting Hung for help with sample preparation, as well as Wright Patterson Air Force Base and Air Force Office of Scientific Research (FA9550- 10-1-0286) for their continued support of this research.

\newpage

\bibliographystyle{osajnl}
\bibliography{PrimaryDatabase}

\end{document}